\documentclass[epj]{webofc}
\usepackage[varg]{txfonts}   
\usepackage{hyperref}

\renewcommand*{\eqref}[1]{Eq.~(\ref{eq:#1})}

\newcommand{\xmax}{X$_{\mathrm{max}}$ }

\wocname{ARENA-2016}
\woctitle{ARENA-2016}
\begin{document}
\title{Ultimate precision in cosmic-ray radio detection --- the SKA}
%
%

\newcommand{\manchester}{School of Physics \& Astronomy, Univ.\ of Manchester, Manchester M13 9PL, United Kingdom}
\newcommand{\nijmegen}{Dept.\ of Astrophysics/IMAPP, Radboud Univ.\ Nijmegen, 6500 GL Nijmegen, The Netherlands}
\newcommand{\astron}{Netherlands Institute for Radio Astronomy (ASTRON), 7990 AA Dwingeloo, The Netherlands}
\newcommand{\karlsruhe}{IKP, Karlsruher Institut f\"ur Technologie, Postfach 3640, 76021 Karlsruhe, Germany}
\newcommand{\erlangen}{ECAP, Univ.\ of Erlangen-Nuremberg, 91058 Erlangen, Germany}
\newcommand{\groningen}{Kernfysisch Versneller Instituut, Univ.\ of Groningen, 9747 AA Groningen, The Netherlands}
\newcommand{\subatech}{Subatech, 4 rue Alfred Kastler, 44307 Nantes cedex 3, France}
\newcommand{\nancay}{Station de radioastronomie de Nan\c cay, Observatoire de Paris, CNRS/INSU, Nan\c cay, France}
\newcommand{\atnf}{CSIRO Astronomy \& Space Science, NSW 2122, Australia}
\newcommand{\karlsruheekp}{EKP, Karlsruher Institut f\"ur Technologie, Kaiserstr. 12, 76131 Karlsruhe, Germany}
\newcommand{\vub}{Astrophysical Institute, Vrije Universiteit Brussel, Pleinlaan 2, 1050 Brussels, Belgium}
\newcommand{\iihe}{Interuniversity Institute for High-Energy, Vrije Universiteit Brussel, Pleinlaan 2, 1050 Brussels, Belgium}
\newcommand{\irvine}{Department of Physics \& Astronomy, University of California, Irvine, CA 92697, USA}
\newcommand{\bologna}{Istituto di Radioastronomy, INAF, via P. Gobetti, Bologna, 4012, Italy}
\newcommand{\curtin}{International Centre for Radio Astronomy Research, Curtin University, Bentley, WA, 6102, Australia}
\newcommand{\mpa}{Max-Planck-Institut für Astrophysik, Karl-Schwarzschildstr. 1, 85748 Garching, Germany}

\author{\firstname{Tim} \lastname{Huege}\inst{1}\fnsep\thanks{\email{tim.huege@kit.edu}}
\and \firstname{Justin D.} \lastname{Bray}\inst{2}
\and \firstname{Stijn} \lastname{Buitink}\inst{3}
\and \firstname{David} \lastname{Butler}\inst{4}
\and \firstname{Richard} \lastname{Dallier}\inst{5,6}
\and \firstname{Ron D.} \lastname{Ekers}\inst{7}
\and \firstname{Torsten} \lastname{En{\ss}lin}\inst{8}
\and \firstname{Heino} \lastname{Falcke}\inst{9,10}
\and \firstname{Andreas} \lastname{Haungs}\inst{1}
\and \firstname{Clancy W.} \lastname{James}\inst{11}
\and \firstname{Lilian} \lastname{Martin}\inst{5,6}
\and \firstname{Pragati} \lastname{Mitra}\inst{3}
\and \firstname{Katharine} \lastname{Mulrey}\inst{3}
\and \firstname{Anna} \lastname{Nelles}\inst{12}
\and \firstname{Beno\^it} \lastname{Revenu}\inst{5}
\and \firstname{Olaf} \lastname{Scholten}\inst{13,14}
\and \firstname{Frank G.} \lastname{Schr\"oder}\inst{1}
\and \firstname{Steven} \lastname{Tingay}\inst{15,16}
\and \firstname{Tobias} \lastname{Winchen}\inst{3}
\and \firstname{Anne} \lastname{Zilles}\inst{4}
}

\institute{\karlsruhe
\and \manchester
\and \vub
\and \karlsruheekp
\and \subatech
\and \nancay
\and \atnf
\and \mpa
\and \nijmegen
\and \astron
\and \erlangen
\and \irvine
\and \groningen
\and \iihe
\and \bologna
\and \curtin
}

\abstract{%
As of 2023, the low-frequency part of the Square Kilometre Array will go online in Australia. It will constitute the largest and most powerful low-frequency radio-astronomical observatory to date, and will facilitate a rich science programme in astronomy and astrophysics. With modest engineering changes, it will also be able to measure cosmic rays via the radio emission from extensive air showers. The extreme antenna density and the homogeneous coverage provided by more than 60,000 antennas within an area of one km$^2$ will push radio detection of cosmic rays in the energy range around 10$^{17}$ eV to ultimate precision, with superior capabilities in the reconstruction of arrival direction, energy, and an expected depth-of-shower-maximum resolution of $<10$~g/cm${^2}$.
}
\maketitle
\section{Introduction}

Over the past decade, radio detection of cosmic-ray air showers has made tremendous progress \citep{HuegePLREP}. Today, radio antenna arrays routinely complement classical cosmic-ray observatories, and have proven to contribute valuable information with competitive resolutions achieved in the determination of the relevant air-shower parameters \citep{SchroederReview}. Arrival direction and energy of the primary particle can be measured accurately even with sparse antenna arrays where individual detector stations are spaced 100 to 200~metres apart. Determination of the depth of shower maximum (X$_{\mathrm{max}}$) is also possible with such sparse arrays, with resolutions of approximately 40~g/cm$^{2}$ achieved with today's analysis approaches \citep{TunkaRexCrossCalibration,GateARENA2016}. The true power for \xmax measurements with radio techniques, however, lies in precision measurements of the radio-emission footprints with dense antenna arrays.

In this article, we discuss the potential of ultra-precise air-shower measurements with the upcoming Square Kilometre Array (SKA)\footnote{Throughout this article, we use the shorthand SKA to refer to the phase~1 implementation of the low-frequency aperture-array part of the SKA, which will measure in the frequency range from 50--350~MHz.}. We illustrate the expected improvement over the existing Low Frequency Array (LOFAR) telescope, describe the science potential and discuss the necessary engineering changes to the SKA baseline design.


\section{Air-shower detection: from LOFAR to the SKA}

In low-frequency radio astronomy, LOFAR can be seen as a pathfinder towards the SKA. The same is true for cosmic-ray detection with dense antenna arrays \citep{LOFARCosmicRays}. Only with the impressive results achieved by LOFAR did the potential for air-shower detection with the SKA become apparent. In particular, the resolution in the determination of \xmax achieved with LOFAR \citep{LOFARNatureXmax} sparked the interest in air-shower detection with the SKA.

LOFAR uses a top-down analysis approach: For each measured air-shower event nearly 100 air-shower simulations with proton- and iron-primaries are generated with the microscopic CoREAS Monte Carlo code \citep{HuegeARENA2012a}. The individual simulations are then compared with the measurements to find the air-shower parameters that yield the best agreement. It turns out that \xmax is the one quantity that governs the agreement, and hence it can be read off directly from the best-fitting simulations. The average \xmax resolution achieved with this analysis techique amounts to 17~g/cm$^{2}$ \citep{LOFARNatureXmax}. The best events even reach an \xmax determination to within better than 10~g/cm$^{2}$. For reasons explained below, we expect that the SKA will yield an \xmax measurement to within 10~g/cm$^{2}$ for each and every measured cosmic-ray shower. This is significantly better than the 20~g/cm$^{2}$ achievable with the most precise techniques available so far, namely fluorescence- and Cherenkov-light detection (see, e.g., \citep{AugerNIM2014}).

\subsection{Increased antenna density and homogeneity}

\begin{figure}[ht]
\centering
\includegraphics[width=0.47\textwidth]{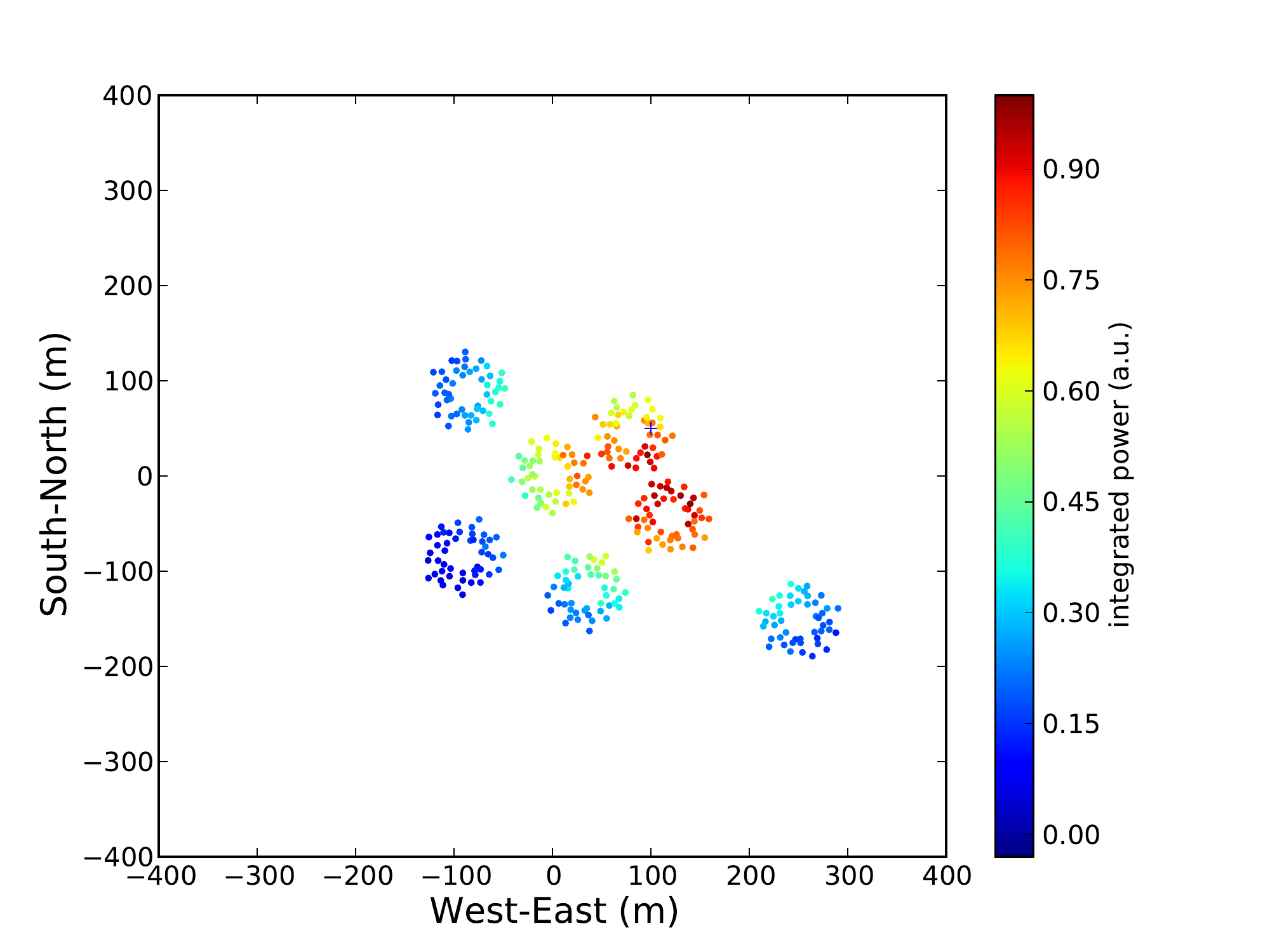}
\includegraphics[width=0.47\textwidth]{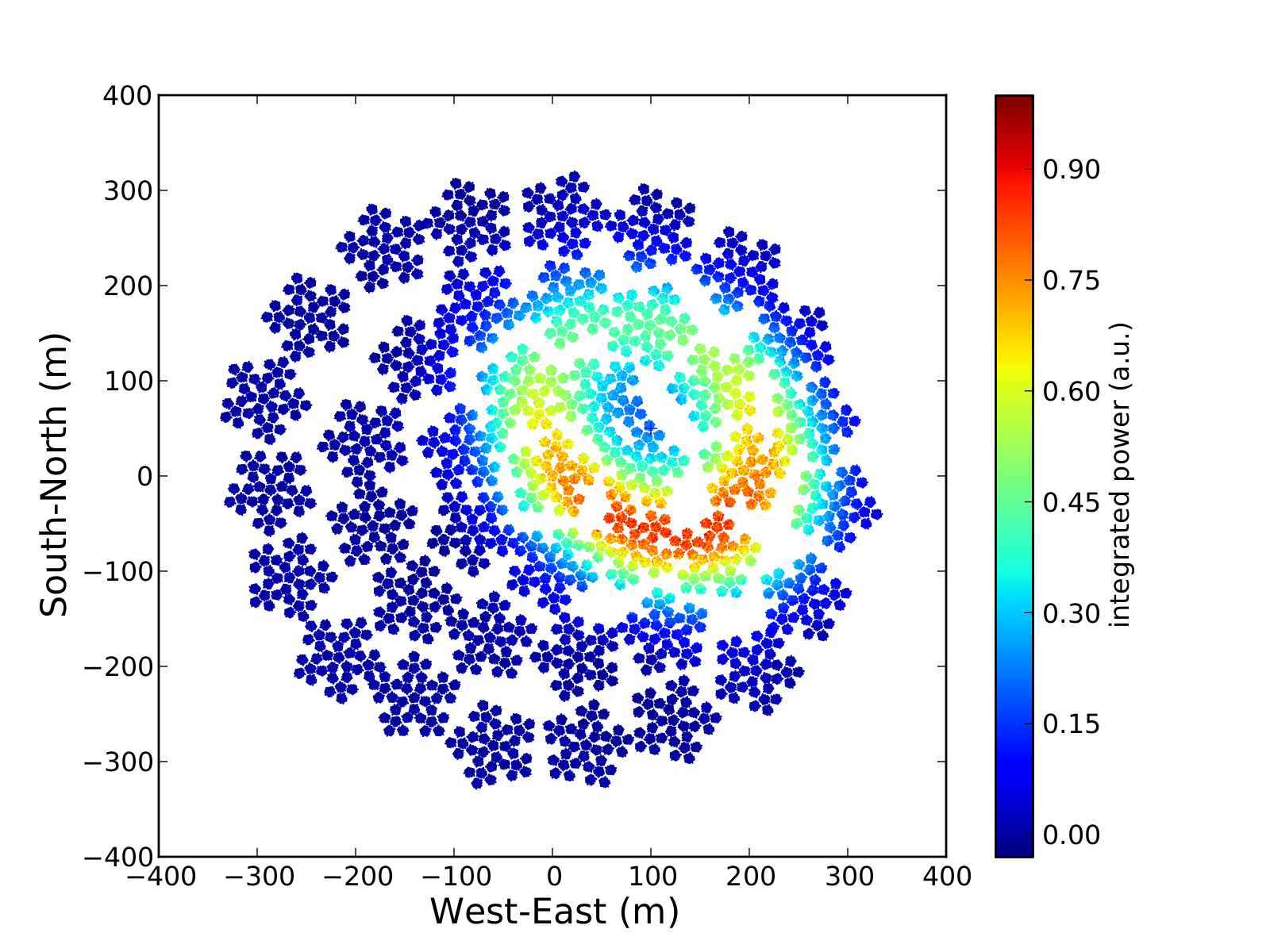}
\includegraphics[width=0.85\textwidth]{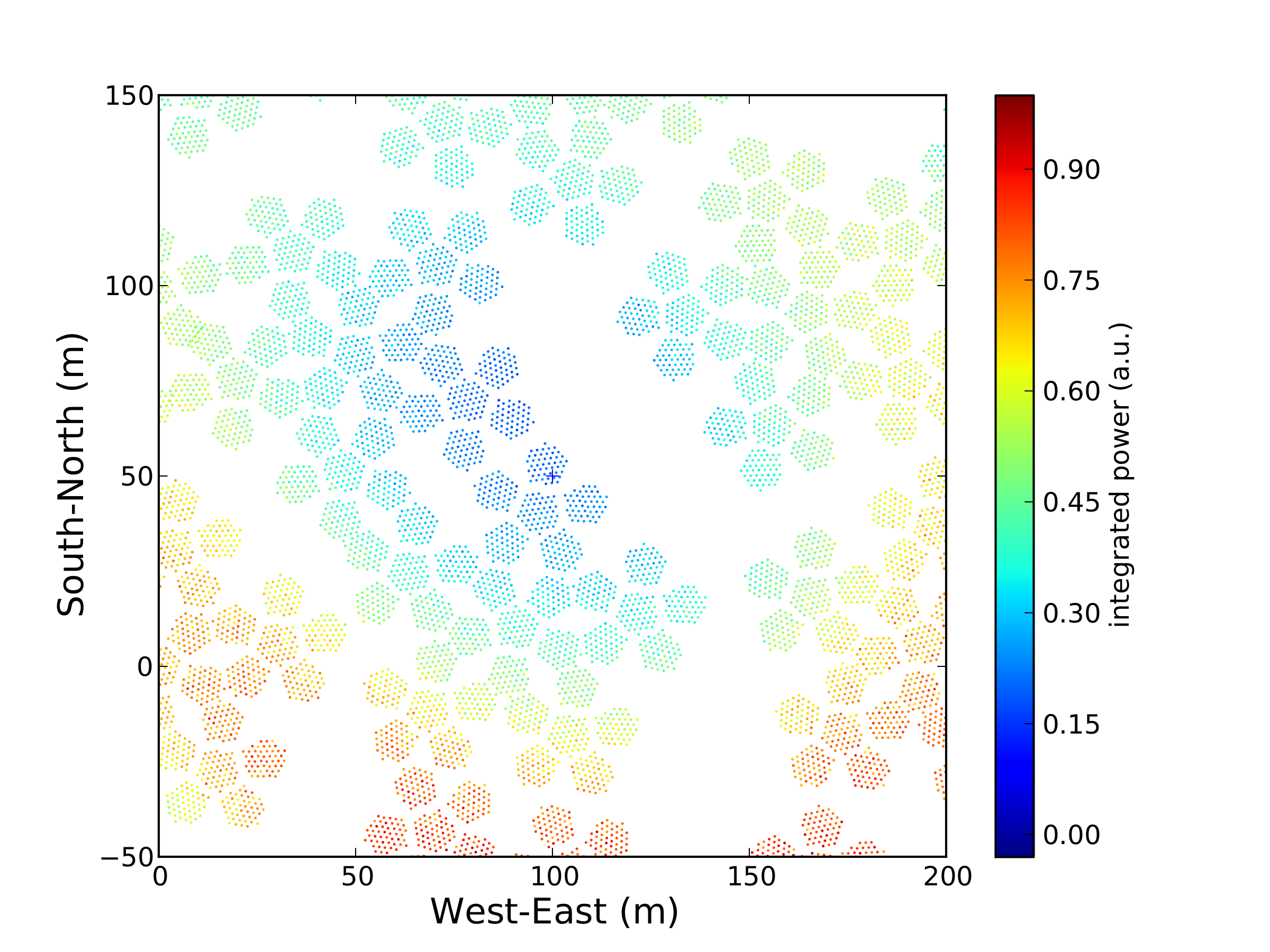}
\caption{CoREAS-based simulation of how the air-shower radio emission would be sampled by LOFAR (top-left) and the SKA core (top-right and zoom-in at bottom). The SKA will measure the radio-emission footprint with an extremely dense and homogeneous array of antennas, yielding superior reconstruction quality on an event-to-event basis. The appearance of a clear ring-structure in the SKA simulation is related to the presence of a Cherenkov ring at higher frequencies. Plots are from the simulation study presented in \citep{ZillesARENA2016SKA}.}
\label{fig:showsim}
\end{figure}

The key factor for the increased measurement precision of the SKA with respect to LOFAR lies in the layout of the antenna array. While LOFAR constitutes a fairly inhomogeneous array with significant gaps in between radio antennas (see Fig.\ \ref{fig:showsim} top-left), the SKA will possess a very dense, very homogeneous core of radio antennas (see Fig.\ \ref{fig:showsim} top-right). The radio-emission footprint has several characteristics sensitive to the depth-of-shower maximum \citep{HuegePLREP}, and for LOFAR only such events where these features are sampled favorably with the inhomogeneous antenna array can yield supreme \xmax resolution. For the SKA, any ``contained'' event will be sampled very precisely (see Fig.\ \ref{fig:showsim} bottom), and will thus be reconstructable with superior precision. Initial simulation studies confirm this expectation with typical \xmax resolutions of $\approx $ 6~g/cm$^{2}$ achieved with a LOFAR-style analysis \citep{ZillesARENA2016SKA}. In combination with experimental uncertainties, a practical resolution of $\approx $ 10~g/cm$^{2}$ thus seems achievable. 

\subsection{Increased frequency coverage}

Also the increased frequency coverage of 50--350~MHz of the SKA as opposed to 30--80~MHz for LOFAR (in the most-used low-frequency observation mode) will benefit the reconstruction quality. Again, initial simulation studies \citep{ZillesARENA2016SKA} illustrate this improvement; a pure-simulation reconstruction for an SKA-like array with a frequency coverage from 30--80~MHz achieves an $\approx$ 10~g/cm$^{2}$ \xmax resolution, while the frequency range from 50--350~MHz yields an \xmax resolution of $\approx$ 6~g/cm$^{2}$.

\subsection{Increased event statistics}

While cosmic-ray detection with LOFAR has provided impressive results, it does suffer from rather limited event statistics. The reasons for this are two-fold: So far, only the innermost area of the LOFAR core can be used for air-shower detection, limiting the fiducial area to less than 0.1~km$^{2}$. Furthermore, cosmic-ray detection is not active 100\% of the time, mostly because technical and organisational hurdles prevent fully commensal operation. This resulted from the fact that cosmic-ray detection capability was not a priority feature from the beginning of the design phase. With the SKA, we aim for 100\% commensal operation and a fiducial area of roughly 1~km$^{2}$, increasing the event statistics by a factor of more than 100 over LOFAR. This will make the energy range from a few times 10$^{16}$~eV up to a few times 10$^{18}$~eV accessible for detailed measurements, with approximately 10,000 events measured per year above 10$^{17}$~eV.


\section{Science potential for air-shower detection with SKA}

The science potential for cosmic-ray detection with the SKA lies in precision measurements. Several scientific goals can be addressed with SKA air-shower measurements.

\subsection{High-precision mass composition measurements}

The energy range accessible with the SKA, from a few times 10$^{16}$~eV up to a few times 10$^{18}$~eV, is suspected to harbour a transition from Galactic to extragalactic cosmic rays \citep{KASCADETransition2013,Blasi2013}. The key metric needed to investigate this hypothesis is highly precise composition information. With supreme \xmax resolution, the SKA will deliver very-high-quality data in this important energy range. Exploiting information beyond a per-event determination of X$_{\mathrm{max}}$, we hope that even a separation between individual particle species, possibly even between proton and Helium, could become feasible. 

\subsection{Particle and air-shower physics}

The air shower evolution, in particular X$_{\mathrm{max}}$, is sensitive to hadronic interaction physics. It has been shown before that the proton-air cross section, multiplicity of secondaries in interactions, elasticity and pion charge ratios can be probed with air-shower measurements \citep{UlrichEngelUnger2011}. With the high-precision measurements provided by the SKA, such studies in the very forward direction and at energies beyond those accessible by the Large Hadron Collider could be carried out with minimised systematic uncertainties. If the deployed particle detector array allows a measurement of the muonic component of the air showers, even more detailed studies of the air-shower physics will become possible.

\subsection{Tomography of the electromagnetic cascade}

The LOFAR analysis approach employed today only uses the integrated power measured by the antennas. In addition, there are measurements of signal timing, pulse shape and polarisation that very likely carry additional information. Furthermore, there is phase information in the radio signal that is currently not taken into account at all. We expect that with near-field interferometric techniques, it should be possible to do ``tomography'' of the electromagnetic cascade in the air shower, and thus provide information that goes well beyond the pure determination of X$_{\mathrm{max}}$.

\subsection{Physics of thunderstorms and lightning}

It has been shown previously that strong electric fields in the atmosphere, in particular those occurring during thunderstorms, influence the radio emission from air showers \citep{BuitinkApelAsch2006}. In fact, properties of the atmospheric electric fields can be probed in situ by detailed analysis of the measured radio emission and comparison with theoretical calculations \citep{LOFARThunderstorms2015}. The air-shower tomography development may lead as well to imaging techniques for the reconstruction of charge flows during lightnings. With the SKA, it will thus be possible to probe the physics of thunderstorms and also test possible connections between lightning initiation and cosmic-ray air showers.


\section{Engineering changes to the baseline SKA-design}

The SKA baseline design does not foresee the capabilities needed for air-shower detection. Here we shortly describe the required engineering changes that have been proposed to the SKA management and are currently under consideration.

\subsection{Antenna buffering}

Buffers are foreseen in the current SKA design to temporarily store beam-formed quantities. For air-shower detection these buffers would need to additionally store the raw signals digitised at individual antennas. With 800~MHz sampling, at least 8-bit, preferably 12-bit dynamic range, and a buffer depth of 10~ms, 1.4~TB of buffers would be needed for 60,000 dual-polarised antennas. This buffering should be carried out with 100\% duty cycle, fully commensal with other observations. On a trigger, 50~$\mu$s worth of data need to be read out, which amounts to 7.2~GB of data per event.

\subsection{Deadtime versus read-out time}

We estimate a readout rate of 1~shower per minute at 10$^{16}$~eV. With the above numbers, the average data rate for readout would be 120~MB/s. This is very small in comparison with the continuous data stream handled by the SKA. However, cosmic rays arrive with a Poissonian time distribution and will thus cause bursts of data that need to be read out quickly. Assuming a data rate of 2.4~GB/s, leading to 3~seconds of readout time during which the buffers are frozen, we will accumulate a tolerable deadtime of 5\%.

\subsection{Trigger generation}

\begin{figure}[htb]
\centering
\includegraphics[width=0.48\textwidth]{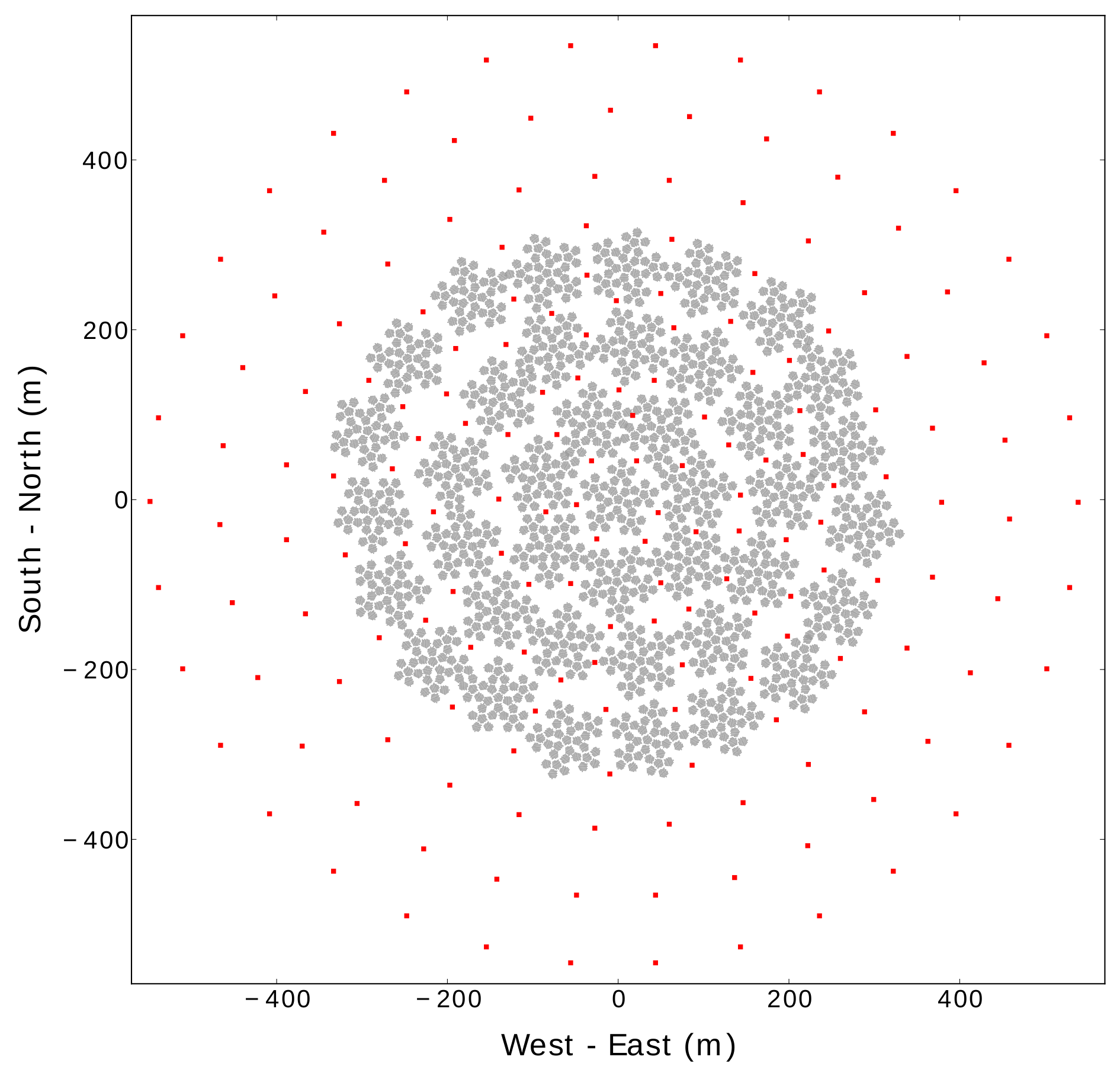}
\caption{Possible layout of a particle detector array with 180 detectors (red squares) in between and outside the antenna stations of the SKA (gray dots).}
\label{fig:pdarray}
\end{figure}

A difficult problem is the generation of an efficient and pure trigger in real time. While it would in principle be possible to generate a trigger from the radio data, we propose to install a particle-detector array for the purpose of triggering. This will also gather valuable information on the particle content of the air shower which can be included in the event reconstruction. The particle detector array should become efficient at 10$^{16}$~eV and thus have an average detector spacing of 50--100~m. The fiducial area should extend beyond the core of antennas because non-contained events will also be usable. An attractive option is to deploy 180~scintillating particle detectors with a size of 3.6~m$^{2}$ each, available from the dismantled KASCADE array \citep{antoni2003}, between the SKA antenna stations (see Fig.\ \ref{fig:pdarray}). The detectors could be read out via the same kind of RF-over-fibre links that are used for the antenna readout. Assuming 4~channels per particle detector, this would require an additional 720~readout channels (on top of the 262,144 needed for the entire antenna array). Special care has to be taken of the radio-quietness of the particle detectors. Experience with LOFAR has shown that careful shielding ensures RFI-quietness.

\subsection{Benefits for the SKA}

The low-level readout capability needed for cosmic-ray measurements will greatly benefit the SKA for diagnostic purposes. Experience with LOFAR shows that many low-level problems (swapped cables, damaged low-noise amplifiers, broken antennas, time synchronisation problems, ...) could be diagnosed with the per-antenna data taken in the cosmic-ray mode. Such problems are otherwise very difficult to track down and will generally deteriorate the quality of astronomical observations.


\section{Conclusion and outlook}

With moderate engineering changes, the SKA can become a cosmic-ray detector that will perform extremely precise measurements of individual air showers in the energy range from a few times 10$^{16}$ to a few times 10$^{18}$~eV. This would enable a rich scientific harvest in return on a very limited additional investment. The SKA focus group on high-energy cosmic particles\footnote{http://astronomers.skatelescope.org/home/focus-groups/high-energy-cosmic-particles/} has proposed the necessary engineering changes to the SKA management, where they are currently under consideration.



%
%
%
%

\end{document}